\documentclass[useAMS,usenatbib]{mn2e}
\usepackage{graphicx,amsmath,times}

\makeatletter

\newcommand{\Rmnum}[1]{\expandafter\@slowromancap\romannumeral #1@}
\makeatother

\title[MHD instabilities in accretion mounds]{MHD instabilities in accretion mounds - \Rmnum{1}: 2D axisymmetric simulations}
\author[Mukherjee, Bhattacharya and Mignone]{Dipanjan Mukherjee$^{1}$\thanks{E-mail:
dipanjan@iucaa.ernet.in}, Dipankar Bhattacharya$^{1}$\footnotemark[1]\thanks{E-mail:
dipankar@iucaa.ernet.in} and Andrea Mignone$^{2}$ \thanks{E-mail: mignone@ph.unito.it }\\
$^{1}$IUCAA, Post Bag 4, Pune, India - 411007\\
$^{2}$Dipartimento di Fisica Generale, Università degli Studi di Torino
, Via Pietro Giuria 1, 10125 Torino, Italy}
\begin{document}

\date{Submitted to MNRAS on 29th June, 2012.}

\pagerange{\pageref{firstpage}--\pageref{lastpage}} \pubyear{2012}

\maketitle

\label{firstpage}

\begin{abstract}
We have performed stability analysis of axisymmetric accretion mounds on neutron stars in High Mass X-ray Binaries (HMXB) by 2-D MHD simulations with the PLUTO MHD code. We find that the mounds are stable with respect to interchange instabilities, but addition of excess mass destabilizes the equilibria. Our simulations confirm that accretion mounds are unstable with respect to MHD instabilities beyond a threshold mass. We investigate both filled and hollow mounds and for the latter also compute the expected profile of cyclotron resonance scattering features (CRSF). In comparison to the CRSF from filled mounds reported in our earlier work, hollow mounds display wider and more complex line profiles.
\end{abstract}


\section{Introduction}
Neutron stars in accreting X-ray pulsars accrete matter from the companion star either from stellar winds \citep{ostriker73} or through disc accretion by Roche lobe overflow \citep{ghosh77,romanova02,romanova03}. They can be broadly classified into two classes: 1) high mass X-ray binaries (HMXB) with companion stars of masses several times the solar mass and neutron stars with high surface magnetic field $\sim 10^{12}$G and  2) low mass X-ray binaries (LMXB) with companion stars of masses less than a solar mass and neutron star magnetic fields several orders lower in magnitude $\sim 10^{7}$G $-10^{9}$G (see \citet{dipankar91} for a review). In this paper we consider the effect of accretion on the evolution of surface magnetic field of HMXB sources by the formation of accretion mounds.

The accreted matter in HMXB passes through a shock, gradually settling down on the polar cap to form an accretion mound. X-ray emission from such mounds show characteristic cyclotron resonance scattering features (CRSF) \citep{harding87,araya99,araya2000,becker07}. The CRSF depends on the magnetic field of the local emitting region, and hence serve as a tool to understand the structure of accretion columns. CRSF often show complex line features and characteristic variations with rotation phase and the luminosity of the neutron star \citep{coburn02,heindl04,mihara07,lutovinov08}. Explaining such features require appropriate modelling of the structure of the accretion column and the effect of accretion induced field distortion from the accretion mound.

Also, several authors propose that diamagnetic screening of the magnetic field can lower the apparent dipole moment of the neutron star \citep{romani90,cumming01,melatos01,choudhuri02,choudhuri04}. Some recent works on magnetic screening by accretion mounds \citep{melatos04,payne07,vigelius08,vigelius09} report that large mounds of mass $\sim 10^{-5} M_{\odot}$ may form on the neutron star, which can then bury the field as the matter spreads on the surface. However several questions regarding the effects of MHD instabilities \citep{litwin01,cumming01} remain to be addressed fully. Magneto-static solutions of accretion mounds have earlier been found by several authors including \citet{hameury83}, \citet{brown98}, \citet{melatos04} and \citet{dipanjan12}. It was shown in \citet{dipanjan12} (hereafter MB12) that magneto-static solutions cannot be found for mounds beyond a threshold height (and mass), which may be indicative of the presence of MHD instabilities. Similar results were also reported in \citet{melatos04} (hereafter PM04) where closed magnetic loops were seen to form beyond a threshold mound mass.

In this paper we attempt to study the stability of the accretion mound by 2D axisymmetric MHD simulations with the PLUTO MHD code \citep{andrea07}. The study of the full set of MHD instabilities in such mounds requires global 3D simulations. However, results from 2D simulations would help to identify modes that grow despite of the restrictive assumption of axisymmetry. This will be a stepping stone to future 3D simulations where many other modes may grow simultaneously. Here we investigate the presence of interchange instabilities as predicted for such mounds by \citet{litwin01}, and also the physical cause of the threshold in mound mass obtained in MB12. To study the latter, we add a small amount of mass to an existing GS solution and dynamically evolve the system to see if it settles to a new equilibrium state. This is carried out for different mound sizes up to the threshold mass, at which one expects MHD instabilities to be triggered if the threshold happens to be due to a physical effect. 

Our approach differs from that of PM04 in various aspects. We consider a cylindrical geometry with strict containment of the accreted matter in the polar cap, while PM04 consider spherical geometry with mass loading on all field lines up to the equator. Also, we consider degenerate non-relativistic Fermi plasma near the polar cap surface instead of the isothermal equation of state  used by PM04. As we consider densities as high as $\sim 10^8 {\rm g \, cm}^{-3}$ inside the mound, a degenerate non-relativistic plasma is more appropriate (see MB12 for a discussion).

Early models of accretion column formed by disc-magnetosphere interaction proposed hollow ring-like accretion column on neutron star poles (\citet{basko76}, \citet{ghosh78} and \citet{ghosh79}). Several authors have used hollow ring-like accretion columns to fit the pulse profiles of HMXBs (e.g. \citet{shakura91}, \citet{leahy91}, \citet{riffert93}). \citet{panchenko94} and \citet{klochkov08_herx1} discuss effects of emission from two disconnected rings to explain shape of observed pulse profiles and nature of cyclotron features in the emission from Her X-1. Following the formalism of pulse profile decomposition developed by \citet{kraus95}, ring-like columns have been inferred for sources like Her X-1 \citep{kraus01}, 4U 1909+07\citep{furst11}, A0535+262 \citep{caballero11} and V 0332+53 \citep{ferrigno11}. Even for LMXB sources, ring like polar cap models are preferred for fitting pulse profiles \citep{poutanen09,kajava11}. We therefore perform a study of the structure and stability of hollow accretion mounds  to compare with results from filled mounds.  We also perform simulations of CRSF emission from hollow mounds, following the method described in MB12.

We structure the paper as follows: in Sec.~\ref{sec.numsetup} we outline the numerical set up involved in the problem. We discuss the solution of the Grad-Shafranov equation to determine the structure of the static mound. We also discuss details of the set up of the MHD simulations with PLUTO. In Sec.~\ref{sec.eqstudy} we discuss the testing of the equilibrium solution with PLUTO. In Sec.~\ref{sec.pert} we discuss the method and results of the perturbation analysis with PLUTO to investigate the stability of the mounds. In Sec.~\ref{sec_holo} we discuss the results of the simulations of hollow mounds and we summarise the results in Sec.~\ref{sec.disc}.

\section{Numerical set up}\label{sec.numsetup}
To test the hydromagnetic stability of the confined mound we first evaluate the equilibrium solution to the Magneto Hydrostatic equations by solving the Grad-Shafranov (hereafter GS) equation. The solution of the GS equation is used as initial condition in PLUTO, where perturbation analysis is performed. In the following section we outline the solution of the GS equation and the set up of the simulation using PLUTO.

\subsection{Equilibrium solution from Grad-Shafranov equation}
For an axisymmetric system, one may write the magnetic field in terms of the flux function in cylindrical coordinates as
\begin{equation}\label{fielddecomp}
\mathbf{B} = \frac{\boldsymbol{\nabla} \psi \times \hat{\boldsymbol{\theta}}}{r} \quad (\mathbf{B}_\theta = 0)
\end{equation}
Using eq.~(\ref{fielddecomp}) in the static Euler equation and using separation of variables in cylindrical coordinates using method of characteristics (as in MB12) we get the GS equation for an adiabatic gas ($p=k_{\rm ad} \rho ^\gamma$)
\begin{equation}\label{GSeq}
\frac{\Delta ^2 \psi}{4\pi r^2} = -\rho g \frac{dZ_0}{d\psi}
\end{equation}
where $g$ is acceleration due to gravity and density is given by the equation 
\begin{equation}\label{rhoeq}
\rho =  \left(\frac{g(\gamma -1)}{\gamma {\rm k}_{\rm ad}}\right)^{\frac{1}{\gamma-1}} \lbrack Z_0 (\psi) - z \rbrack ^{\frac{1}{\gamma-1}}
\end{equation}
$Z_0(\psi)$ is the mound height function which determines the shape of the mound. For our work we use the equation of state  for a degenerate non-relativistic zero temperature Fermi plasma with  $\mu _e =2$: 
\begin{equation}
\left. \begin{aligned}  p &=  [(3 \pi ^2)^{2/3}\frac{\hbar ^2}{5m_e}]\left(\frac{\rho}{\mu _e m_p} \right) ^{5/3}\, \\
& = 3.122 \times 10^{22}  \left(\frac{\rho}{10^6 \mbox{ g cm}^{-3}}\right)^{5/3} \mbox{ dynes cm}^{-2} \, \label{eos}
\end{aligned}
\right \}
\end{equation}
Most of the mound will be dominated by degeneracy pressure except for a thin layer at the top ($\sim 4$cm at 1keV plasma, see MB12 for a discussion). Thus effects of thermal stratification would play a limited role, and the zero temperature degenerate equation of state would be an adequate assumption. We solve the  GS equation for an accretion mound of radius $R_p = 1$km, on the poles of a slowly spinning neutron star of mass $1.4 M_\odot$ and radius $R=10$km. The intrinsic field is assumed to be dipolar, which in the polar cap region can be approximated as an uniform field along $\hat{\mathbf{z}}$ ($\mathbf{B}_{\rm p}=B_0 \hat{\mathbf{z}}$). We consider Newtonian gravity with constant acceleration:
\begin{equation}
\mathbf{g} = -1.86 \times 10^{14} \left(\frac{M_*}{1.4M_\odot}\right) \left(\frac{R_s}{10 \rm km}\right)^{-2} \mbox{ cm s}^{-2} \mbox{\hspace{1mm}}\hat{\mathbf{z}}
 \end{equation}
Our set up is similar to that in \citet{hameury83} and \citet{litwin01}. Following MB12, we carry out most of our analysis for the mound height profile:
\begin{equation}\label{parabolicpro}
Z_0(\psi ) = Z_{\rm c}\left(1-\left(\frac{\psi}{\psi _p}\right)^2\right) 
\end{equation} 
where $Z_c$ is the central height of the mound and $\psi _p=(1/2)B_0R^2_p$. This is a smoothly varying parabolic profile in $\psi$ which describes a filled axisymmetric mound. We also discuss the GS solution for a hollow mound  in Sec.~\ref{sec_holo}, which is specified by the mound height function:
\begin{equation}\label{holopro}
Z_0(\psi ) = \frac{Z_{\rm c}}{0.25}\left(0.25-\left(\frac{\psi}{\psi _p} - 0.5\right)^2\right)
\end{equation}
The GS is a coupled non-linear elliptic partial differential equation. We have solved the GS equation by an iterative under-relaxation algorithm with an inner Successive Over-relaxation loop with Chebyshev acceleration \citep{press} as is outlined in MB12. For a given polar magnetic field ($B_p$), the solutions to the GS equations are obtained up to a threshold height $Z_{\rm max}$, beyond which the numerical scheme does not converge to give an unique solution. Details of the numerical algorithm and convergence of the GS solutions have already been discussed in MB12.

\subsection{PLUTO setup: Initialisation }\label{sec.init}
We use the Godunov scheme based MHD code PLUTO \citep{andrea07} to test the stability of the confined mound. The solutions of the GS equation are used as initial condition in PLUTO.  The GS solutions are imported into PLUTO using bi-linear interpolation.  We use the MHD module of PLUTO to solve the full set of ideal magneto hydrodynamic equations: 
\begin{eqnarray}
\frac{\partial \rho}{\partial t} + \boldsymbol{v}\cdot\nabla\rho + \rho\nabla\cdot\boldsymbol{v} &=& 0 \label{eqrho} \\
\frac{\partial\boldsymbol{v}}{\partial t} + \boldsymbol{v}\cdot\nabla\boldsymbol{v} + \frac{1}{\rho}\boldsymbol{B}\times(\nabla\times\boldsymbol{B}) + \frac{1}{\rho}\nabla p &=& g \\
\frac{\partial\boldsymbol{B}}{\partial t} + \nabla\times(\boldsymbol{v}\times\boldsymbol{B}) &=& 0 \label{induceq} \\ 
\frac{\partial p}{\partial t} + \boldsymbol{v}\cdot\nabla p + \rho c^2_s\nabla\cdot v &=& 0 \label{energyeq}
\end{eqnarray}
where the factor $1/\sqrt{4\pi}$ is absorbed in the definition of magnetic field and $c^2_s$ is the speed of sound (which for adiabatic gas is $c^2_s=\gamma p/\rho$). The system is closed by an equation of state (hereafter EOS) which we choose to be either adiabatic ($\rho\epsilon = p/(\gamma-1)$) or barotropic for which $p = p(\rho)$. In the second case eq.~(\ref{energyeq}) is redundant. To investigate the effects of pressure driven interchange modes and gravity driven modes, we perform perturbation analysis with the adiabatic EOS (see Sec.~\ref{sec.zeromean} and Sec.~\ref{sec.massadd}). PLUTO initialisation and boundary conditions are provided in terms of primitive variables $(\rho,\mathbf{v},p,\mathbf{B})$ defined in eq.~(\ref{eqrho}) $-$ eq.~(\ref{energyeq}). The computation is carried out in conservative variables $(\rho,\rho \mathbf{v},E,\mathbf{B})$, where $E= \rho \epsilon + \rho \mathbf{v}^2/2 + \mathbf{B}^2/2$ is the total energy density.

We use the extended generalised Lagrangian multiplier (EGLM) scheme (\citet{mignone10_1}, \citet{mignone10_2}) to preserve the $\nabla\cdot\boldsymbol{B}=0$ constraint. The EGLM scheme preserves the divergence criterion by modifying the induction equation (eq.~\ref{induceq})  with a scalar field function $\psi _{\rm GLM}$ \citep{dedner02} and also the energy momentum equations with extra source terms.  This scheme transports the non-zero divergence errors to the boundary of the domain at the fastest possible characteristic speed, and damp them at the same time. 

For our problem, we have found that the HLL Riemann solver \citep{toro08}, HLLD Riemann solver \citep{miyoshi05} and TVD Lax-Friedrichs solver \citep{toro08} combined with EGLM scheme provide solutions free from numerical instabilities. Due to the presence of very sharp gradients in the physical quantities, higher order schemes need to be employed to reduce numerical errors. A third order Runge-Kutta scheme is used for time evolution and a third order accurate piece-wise parabolic interpolation scheme (PPM scheme as in \citet{colella84}) has been employed.

The simulations were set up using square cells ($\Delta r \simeq \Delta z$) to minimise numerical errors. The resolutions used were less than $\sim 0.5$m as listed in Table.~\ref{tableres} for some sample runs.
\begin{table}
\centering
\begin{tabular}{|l|l|l|l|}
\hline
$Z_c$ & $B_p$ & $N_r \times N_z$ & $\Delta l$ ($\Delta z \simeq \Delta r$) \\
\hline
72m & $10^{12}$G & $1024\times144$ & $\sim 0.43$m \\
65m & $10^{12}$G & $1088\times104$ & $\sim 0.46$m \\
55m & $10^{12}$G & $1272\times88$  & $\sim 0.39$m \\
50m & $10^{12}$G & $1024\times80$  & $\sim 0.43$m \\
25m & $10^{11}$G & $1920\times72$  & $\sim 0.2$m \\
\hline
\end{tabular}
\caption{Sample resolutions for simulation runs.}
\label{tableres}
\end{table}
The physical variables in PLUTO are scaled to non-dimensional forms before initialisation. For example for mounds with polar magnetic field $B_p = 10^{12}$G, we use $\rho = 10^6{\rm g \, cm}^{-3}$ as the density unit, $L_0 = 10^5$cm as the length unit, $B_0=10^{12}$G as the magnetic field unit and $V_{A0} = B_0/\sqrt{4 \pi \rho} = 2.82 \times 10^8 {\rm cm \, s}^{-1}$ as the velocity unit. In these units, time is measured in units of $t_{A} = L_0/V_{A0} = 3.55 \times 10^{-4}$s, which can be taken as the mean Alfv\'en time, while the scale velocity is the mean Alfv\'en velocity. An unique Alfv\'en velocity cannot be prescribed for the whole domain as the Alfv\'en speeds will vary over the domain depending on local density and magnetic field.

\subsection{Boundary Conditions}\label{sec.bc}
For stability studies, we run the simulations with either fixed boundaries where quantities are kept fixed to initial values ($Q=Q_0$) or fixed gradients where the initial gradients are preserved. The fixed gradient boundary implies outflow of perturbed quantities as gradients of perturbations are set to zero ($\nabla Q = \nabla Q_0 + \nabla \tilde{Q} \rightarrow \nabla Q_0, \nabla \tilde{Q} = 0$). The standard outflow boundary condition ($\nabla Q = 0$) is inapplicable for our problem as the initial solution has non-zero gradients at the boundaries of the domain. The fixed gradient boundary condition is applied to the upper and the rightmost boundary. For filled mounds, the inner left boundary is kept fixed as it is close to or equal to the axis of the column. For hollow mounds, the inner left boundary is kept at a fixed gradient to allow for inward flow of perturbed matter. The bottom boundary is kept fixed to simulate a hard crust. The set-up with fixed gradients on the outer sides and fixed crust gives numerically stable solutions, as tested from simulations of the equilibrium solutions obtained from GS-solver (see Sec.~\ref{sec.eqstudy}).

\section{Equilibrium studies}\label{sec.eqstudy}
\begin{figure}
	\centering
	\includegraphics[width = 8cm, height = 8cm,keepaspectratio] {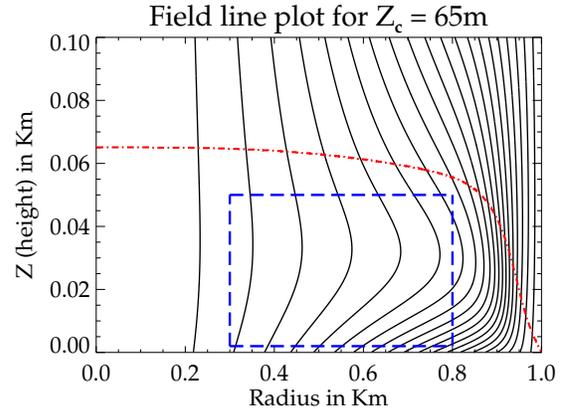}
	\caption{ \small Field lines for a mound of height $Z_c = 65$m with polar unloaded field $B_p = 10^{12}$G. The dash-dotted line in red denotes the top of the mound beyond which density is zero. The total mass of the mound is $\sim 1.63\times10^{-12} M_\odot$.  The dashed blue box in the middle is the PLUTO computation domain, chosen to keep Alfv\'en velocities non-relativistic.  The range of density is $\sim 2.1\times10^6 - 6.7\times 10^6 \mbox{ g cm}^{-2}$ at the top of the mound and  $\sim 3.02\times10^7 - 5.7\times10^7 \mbox{ g cm}^{-2}$. at the bottom.}\label{GS65m}
\end{figure}
The GS solutions for adiabatic mounds have density profiles which go to zero beyond $Z_0(\psi)$ (see eq.~(\ref{rhoeq})). To avoid unrealistic Alfv\'en velocities, we restrict the computation domain inside the mound such that Alfv\'en speeds in the mound are non-relativistic. A typical computation domain is depicted in Fig.~\ref{GS65m} for a mound of height $Z_c =65$m.  We first evolve the initial equilibrium solution without applying perturbation in order to check the stability of the numerical schemes and also to study the effects of initial transients contributed by the numerical errors accumulated in interpolating the solution from GS grid to PLUTO domain.

The solutions have been evolved to $t\sim 80 t_{A}$ for different choices of schemes.  For the set of schemes outlined in Sec.~\ref{sec.init} and Sec.~\ref{sec.bc}, the equilibrium solution remains intact, with very small build up of internal flow velocities. For example, for a mound of height $Z_c = 72$m, at $t \sim 80 t_{A}$, the maximum velocity is $\sim 7.5 \times 10^{-4}$ in normalised units ($\sim 2.15 \times 10^5 {\rm cm \, s}^{-1}$, which is much smaller than typical scale velocities). This shows that the schemes used are free from artificial numerical effects and also verifies the validity of the equilibrium solution obtained from the GS solver. 
\begin{figure}
	\centering
	\includegraphics[width = 6cm, height = 6cm,keepaspectratio] {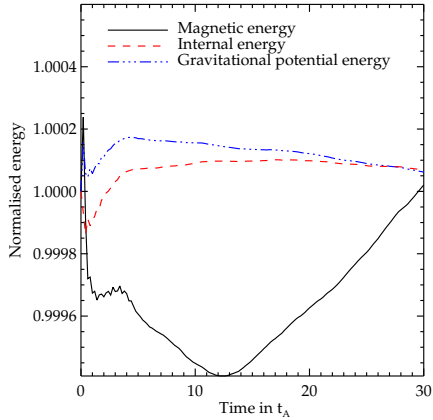}
	\caption{ \small Energy components for zero-mean random perturbation run, normalised to their initial value. Magnetic energy is normalised to $3.7\times10^{22}$, internal energy to $8.9\times10^{23}$ ergs and gravitational potential energy to $6.7\times10^{23}$ ergs. The internal and gravitational energy components remain almost constant ($\sim 0.02\%$ change from initial value). The magnetic energy initially decreases as the pockets of perturbed matter settle down, eventually returning to its initial value. This indicates that the system is stable, and when perturbed, settles to a energy state close to the original equilibrium value. }\label{65mzeromean_en}
\end{figure}
\begin{figure}
	\centering
	\includegraphics[width = 6cm, height = 6cm,keepaspectratio] {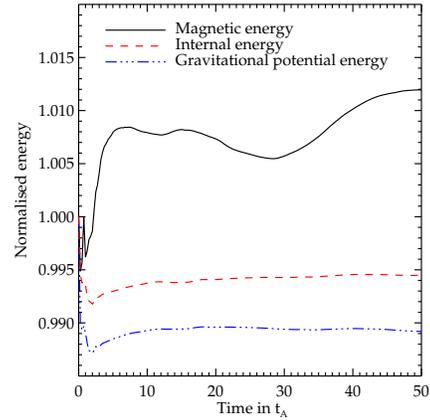}
	\caption{ \small Energy components for random positive perturbation run ($\eta = 5\%$), normalised to their initial value. Magnetic energy is normalised to $5.2 \times 10^{22}$ ergs, internal energy to $9.9\times 10^{23}$ ergs and gravitational potential energy to $7.5\times10^{23}$ ergs. The initial energy is dominated by internal and gravitational energy. The gravitational and internal energy decrease as the system move to a lower energy state following the perturbation. The magnetic energy is seen to increase due to stretching of field lines due to internal flows. }\label{65m_5_en}
\end{figure}

\begin{figure*}
	\centering
        \includegraphics[width = 12.0cm, height = 6cm,keepaspectratio] {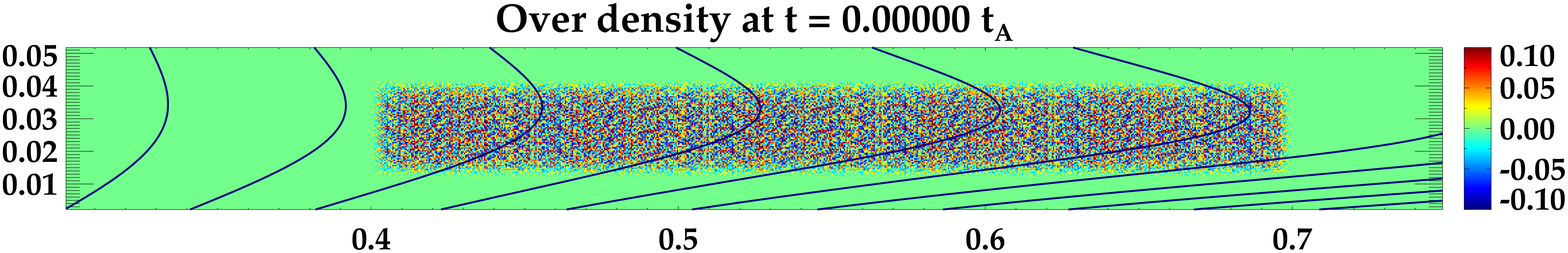}
	\includegraphics[width = 12.0cm, height = 6cm,keepaspectratio] {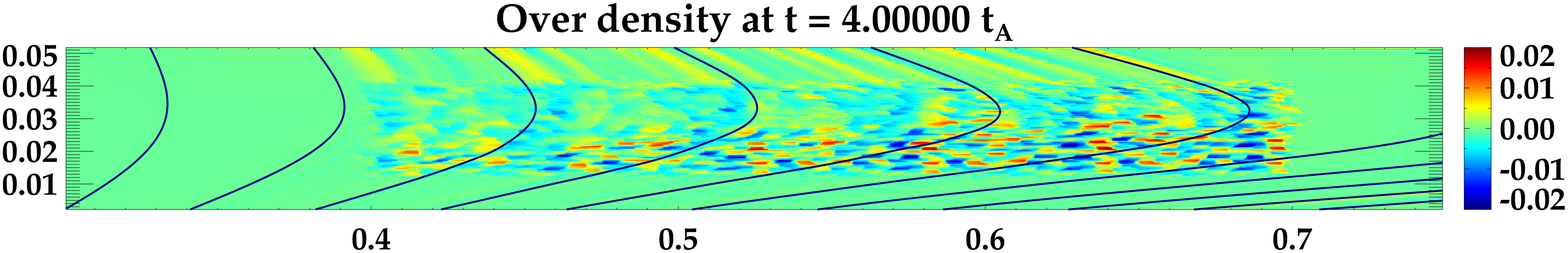}
	\caption{ \small Over-density: $(\rho - \rho _{\rm eq})/\rho_{\rm eq}$ for zero-mean perturbation runs for a mound of height $Z_c = 65$m, polar magnetic field $B_p = 10^{12}$G and perturbation strength $\eta = 10\%$. $\rho_{\rm eq}$ is the unperturbed density from the equilibrium solution. The vertical axis is the height above neutron star surface in kilometres. The horizontal axis is the radius (cylindrical geometry) in kilometres. The PLUTO simulation was carried out with a grid of size $1024 \times 120$. Random perturbation is provided within a rectangular box inside the domain, away from the boundaries. The edges of the perturbation region are smoothed exponentially. The perturbation slowly weakens and relaxes into stable pockets of perturbed density by $t \sim 4 t_{\rm A}$ (bottom panel). The magnetic field lines are plotted in black.}\label{zeromean}
\end{figure*}

\section{Perturbation Analysis}\label{sec.pert}
We perturb the equilibrium solution by adding a normalised perturbation field $\xi(r,z)$ to any of the physical quantities 
\begin{equation}\label{perteq}
Q=Q_0(1+\eta \xi(r,z))
\end{equation}
where $\eta$ is a positive number signifying the perturbation strength. The perturbations are kept away from the boundaries on all sides. This is to preserve the equilibrium at the boundary layers and avoid spurious interaction with the boundary. For our studies we apply a random perturbation on the density inside the simulation domain, namely $\xi$ is assigned a random value at each grid point within the perturbation zone. The edges of the perturbing region are smoothed with an exponential function to avoid sharp gradients which can lead to spurious effects. The lack of any preferred perturbation scale should allow the growth of the fastest growing modes. The perturbation analysis is performed for mounds of different heights up to the threshold height $Z_{\rm max}$ beyond which the GS-solver does not converge, as has been found in MB12. 

\subsection{Zero-mean perturbations: interchange modes}\label{sec.zeromean}
\begin{figure*}
	\centering
	\includegraphics[width = 12cm, height = 6cm,keepaspectratio] {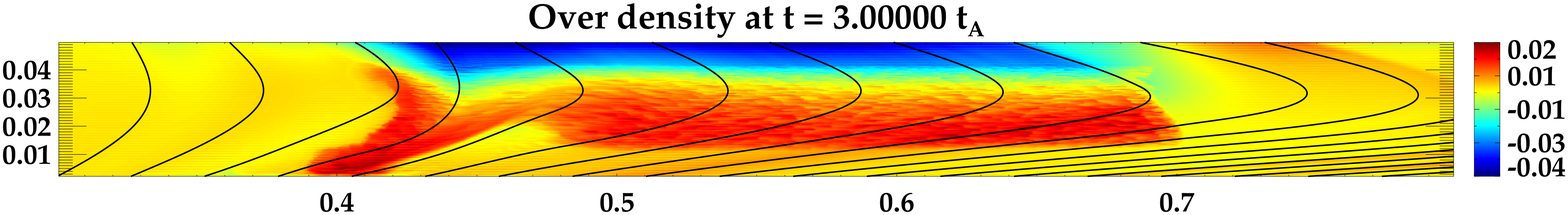}
	\includegraphics[width = 12cm, height = 6cm,keepaspectratio] {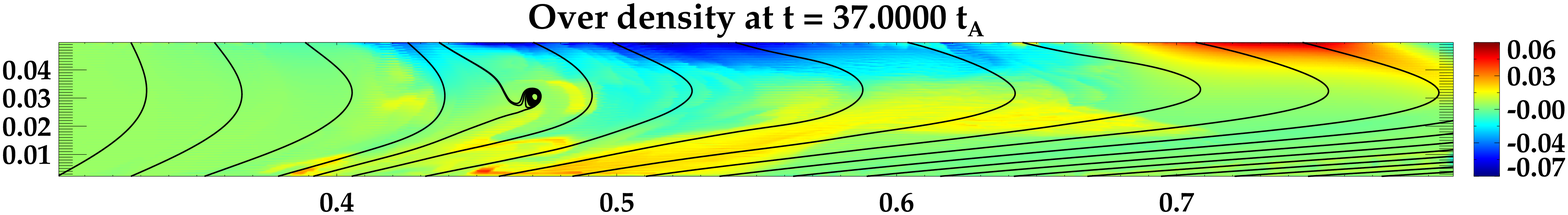}
	\includegraphics[width = 12cm, height = 6cm,keepaspectratio] {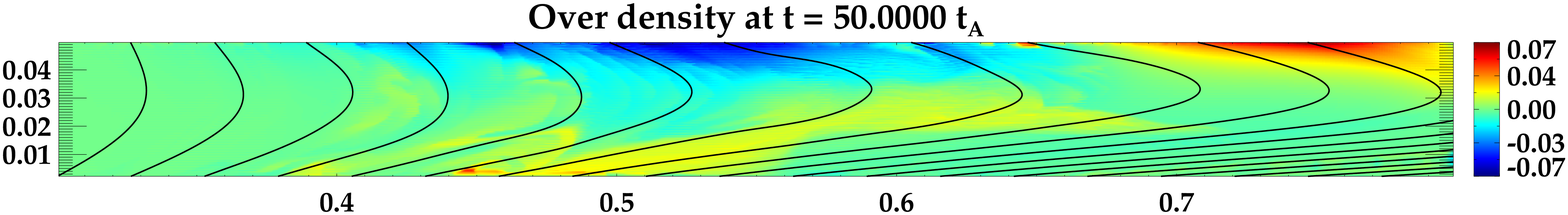}
	\caption{ \small Over-density: $(\rho - \rho _{\rm eq})/\rho_{\rm eq}$ at different times for a positive density perturbation with strength $\eta = 3\%$ in a mound of height $Z_c = 65$m and polar magnetic field $B_p \sim 10^{12}$G. The simulation was carried out with a grid of size $1088 \times 104$. Horizontal and vertical axes are the same as in Fig.~\ref{zeromean}. The perturbations result in the formation of closed loops but the solution eventually settles down to a steady state.}\label{65m_3_rho}
\end{figure*}
\begin{figure*}
	\centering
	\includegraphics[width = 12cm, height = 6cm,keepaspectratio] {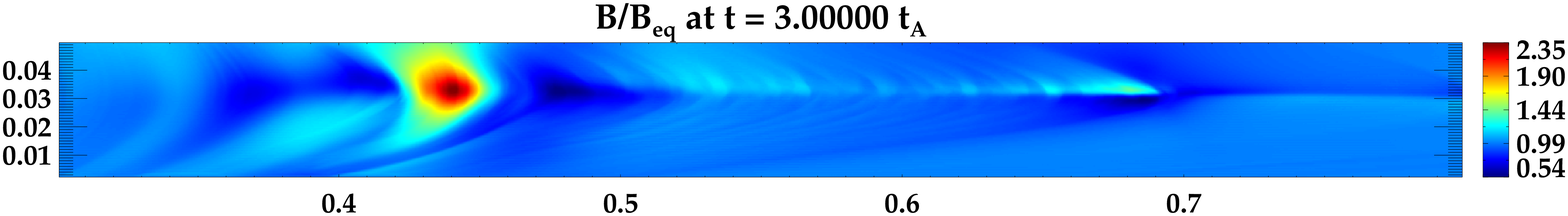}
	\includegraphics[width = 12cm, height = 6cm,keepaspectratio] {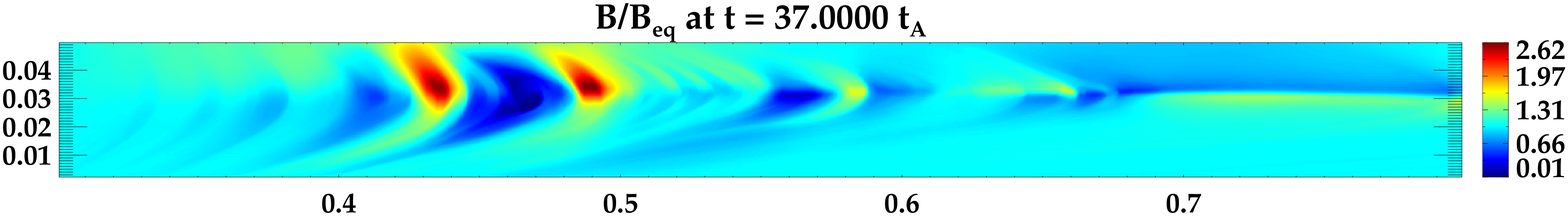}
	\includegraphics[width = 12cm, height = 6cm,keepaspectratio] {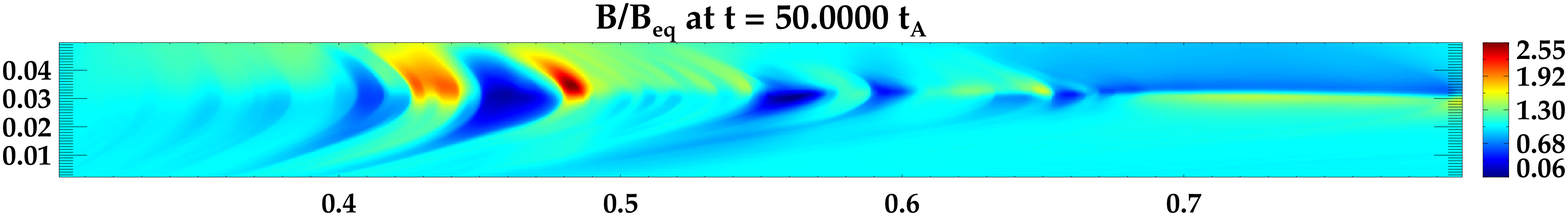}
	\caption{ \small Magnetic field magnitude normalised to the local equilibrium value for the simulation described in Fig.~\ref{65m_3_rho}. Bunching of field lines forms pockets of excess field over equilibrium value, which eventually get smeared and start to dissipate.}\label{65m_3_bmag}
\end{figure*}
Zero-mean random perturbation with $\langle \xi \rangle = 0$, implies rearranging of density from the equilibrium solution without adding any net mass. In this case, the system quickly converges to stable pockets of perturbations, irrespective of perturbation strength ($\eta$ in eq.~\ref{perteq}). See Fig.~\ref{zeromean} for the results of a run with perturbation strength $\eta = 10\%$. The system settles down to an energy state close to the original equilibrium value (see Fig.~\ref{65mzeromean_en}). However, for larger perturbation strengths, a longer time is taken to relax into stable pockets of perturbed matter. For example a mound with $B_p = 10^{12}$G and $Z_c = 65$m stabilizes after $t \sim 1 t_{A}$ for $\eta = 2\%$ and $t\sim 4 t_A$ for  $\eta = 10\%$. 

The perturbation tests have been carried out for mounds of different heights and polar magnetic field strengths. No instabilities are seen at the threshold mound heights, e.g. $Z_c \sim 72$m for $B = 10^{12}$G and $Z_c \sim 25$m for $B = 10^{11}$G etc.  The simulations show that the mounds are stable with respect to small departures from  equilibrium resulting from rearrangement of flux tubes. Thus interchange or ballooning modes are not seen in 2-D axisymmetric simulations of the mounds.

\begin{figure*}
	\centering
	\includegraphics[width = 12cm, height = 6cm,keepaspectratio] {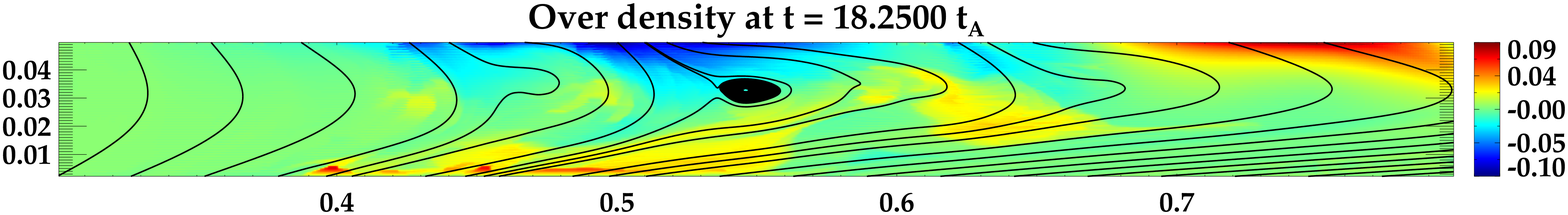}
	\includegraphics[width = 12cm, height = 6cm,keepaspectratio] {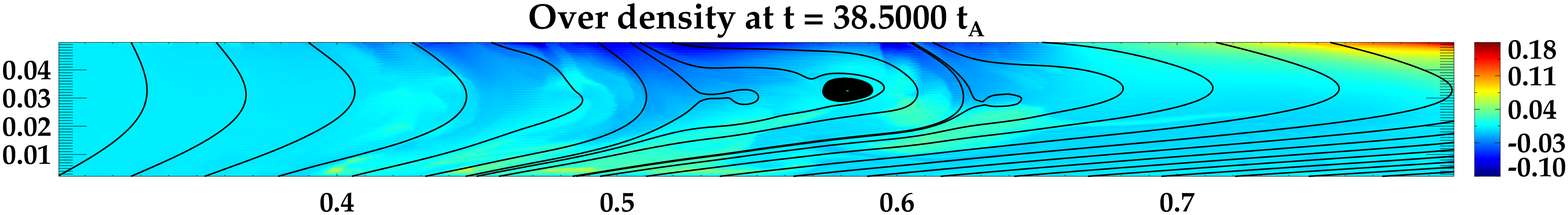}
	\includegraphics[width = 12cm, height = 6cm,keepaspectratio] {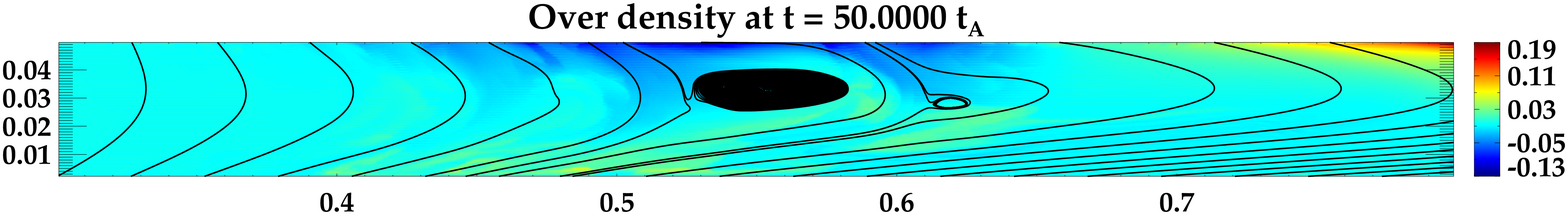}
	\caption{ \small Over-density  $(\rho - \rho _{\rm eq})/\rho_{\rm eq}$ at different times for a positive density perturbation with strength $\eta = 5\%$ in a mound of height $Z_c = 65$m and $B_p = 10^{12}$G. The simulation was carried out with a grid of size $1088 \times 104$. Horizontal and vertical axes are the same as in Fig.~\ref{zeromean}. Reconnection of field lines forms closed loops at multiple sites. The system does not relax to any steady state solution within the duration of the run. The closed loops grow with time indicating the onset of unstable modes. }\label{65m_5_rho}
\end{figure*}
\begin{figure*}
	\centering
	\includegraphics[width = 12cm, height = 6cm,keepaspectratio] {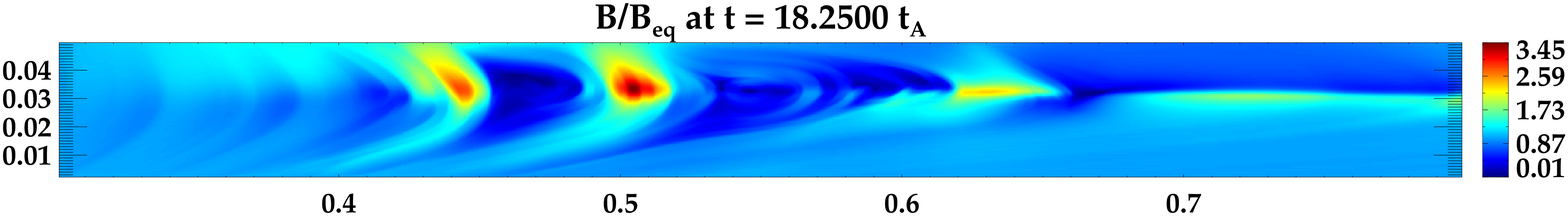}
	\includegraphics[width = 12cm, height = 6cm,keepaspectratio] {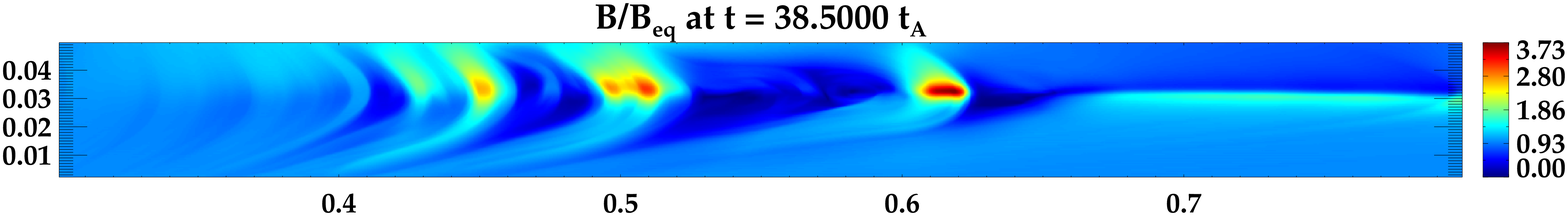}
	\includegraphics[width = 12cm, height = 6cm,keepaspectratio] {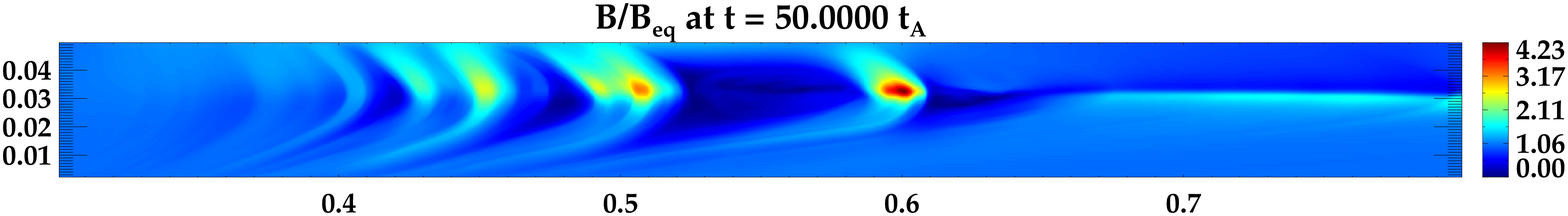}
	\caption{ \small Magnetic field magnitude normalised to the local equilibrium value for the simulation described in Fig.~\ref{65m_5_rho}.  Bunching of field lines cause pockets of excess field over equilibrium value which do not settle to any steady state.}\label{65m_5_bmag}
\end{figure*}

\subsection{Adding excess mass to equilibrium solution}\label{sec.massadd}
In order to study the effect on the mound of the addition of matter which eventually descends due to gravity,  we apply a positive definite random perturbation field: $\langle \xi \rangle  > 0$ on the density without any corresponding change in pressure.  Such a change in density implies local departure of $k_{\rm ad}$ from that in eq.~\ref{eos}. In this work we do not attempt to model the exact composition of the accretion mound. The perturbations were set up to ensure that the added matter is heavier than its surroundings and will descend due to gravity, thus triggering the gravity driven modes. However, a change in $k_{\rm ad}$ can indeed occur due to changes in chemical composition e.g. $\eta \sim 5\%$ local perturbation on a $Z_c \sim 65$m mound would correspond to a change of mean molecular weight by $\Delta \mu _e \sim 0.1$.

 The  added mass settles down along the field lines, dragging and distorting the equilibrium field configuration in the process. For small perturbation strengths ($\eta \simeq 1\%$ for mound of height $Z_c = 65$m) the matter quickly settles down to a new equilibrium, without appreciable distortion of the field lines. With an increase in $\eta$ beyond a threshold, e.g. $\eta _T \sim 3\%$ for $Z_c = 65$m and $B_p = 10^{12}$G mound, magnetic Rayleigh-Taylor type instabilities are triggered by the descending heavier matter and results in the formation of closed loops due to reconnection of field lines (see Fig.~\ref{65m_3_rho}).\footnote{Note that although the simulation is ideal MHD, numerical resistivity allows dissipation and reconnection to occur.} Bunching of field takes place in the radial direction (e.g. Fig.~\ref{65m_3_bmag}) and the system eventually relaxes to a steady state. 

Further increase in perturbation strength, e.g. $\eta \sim 5 \%$ for $Z_c = 65$m, disrupts the equilibria completely. Several closed loops are formed across the perturbed region (see Fig~\ref{65m_5_rho} and Fig.~\ref{65m_5_bmag}). Individual closed loops merge to form larger knots without showing any signs of decay. From Fig.~\ref{65m_5_en} we see that gravitational potential energy and internal energy decreases from initial value, whereas magnetic energy increases with time. This indicates that internal flows stretch and twist the field lines converting internal energy and gravitational energy to magnetic energy. The system does not relax to a steady state within the run time of the simulation ($t\sim 50 t_A$). Thus for a mound with $Z_c = 65$m and $B_p = 10^{12}$G, the threshold perturbation strength is $\eta_T \sim 3\%$ beyond which gravity and pressure driven modes disrupt the MHD equilibria.  

Convergence has been tested by running the simulations for successive higher resolutions: .e.g. for $Z_c = 65$m, $B_p = 10^{12}$G with positive random perturbation of strength $\eta = 5\%$ simulations were carried out for resolutions ($1088\times104$), ($2176\times208$) and ($4352\times416$). It was seen that MHD instabilities persist on increase of resolution. Increase in resolution reduces numerical resistivity, thus decreasing cross field diffusion. The field lines are then more prone to be deformed by gravity driven modes triggered by the weight of the overlying matter.

With an increase in mound height, it is easier to excite such unstable behaviour. The threshold perturbation strength is larger for mounds of smaller height: for $Z_c = 45$m and $B_p = 10^{12}$G, $\eta_T \sim 7\%$. Mounds near the GS threshold height $Z_{\rm max}$ ($ \sim 72$m for $B_p = 10^{12}$G; $ \sim 25$m for $B_p = 10^{11}$G) are only marginally stable at $\eta_T \simeq 1\%$.  Thus, mounds higher than a threshold (as previously obtained in MB12)  are prone to gravity driven Rayleigh-Taylor and pressure driven instabilities on addition of excess mass, and stable magneto-static solutions cannot be obtained.

\section{Hollow mound}\label{sec_holo}
\subsection{Grad-Shafranov for hollow mounds}
\begin{figure}
	\centering
	\includegraphics[width = 8cm, height = 8cm,keepaspectratio] {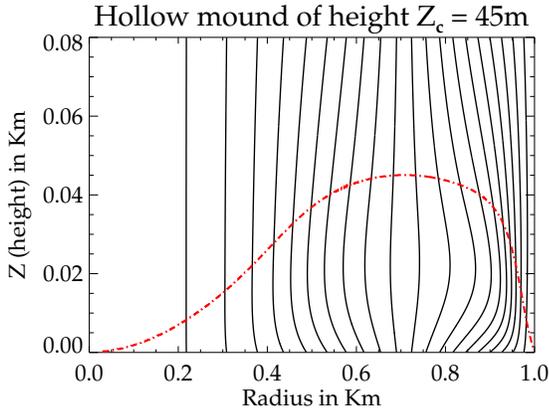}
	\caption{ \small The field lines from GS solution for a hollow mound with mound height function given by eq.~(\ref{holopro}), $Z_c = 45$m and $B_p = 10^{12}$G. The maximum height $Z_c$ occurs at $\sim 698$m from the axis. The red-dashed line represents the top of the mound.}\label{holoGS}
\end{figure}
\begin{figure}
	\centering
	\includegraphics[width = 7cm, height = 7cm,keepaspectratio] {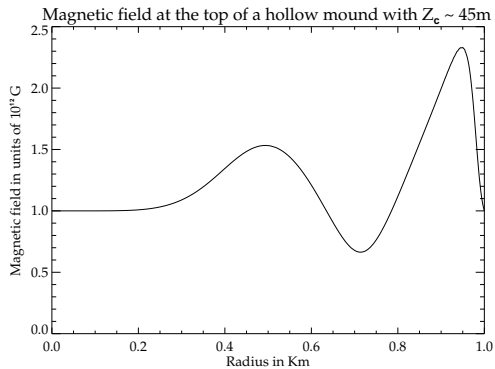}
	\caption{ \small The magnetic field at the top of the hollow mound in Fig.~\ref{holoGS}. Field lines are pushed on either side of the apex ($r \sim 698$m) of the mound resulting in decrease in field at the apex and increase in field strength on either side. Starting from a polar magnetic field strength $B_p = 10^{12}$G, from our GS solution we get minimum field at the top  $\sim 6.63 \times 10^{11}$G and maximum field of $\sim 2.33 \times 10^{12}$G. }\label{holofieldtop}
\end{figure}

For systems with magnetospheric accretion,  mass loading at the accretion disc takes place over a finite range of accretion disk radii ($\Delta_r \sim 0.03 R_A$, $R_A \equiv$ Alfv\'en radius e.g. \citet{ghosh78,ghosh79}). The inner edge of the polar cap ring \footnote{which corresponds to the outermost radius in the accretion disc $\sim R_A+\Delta_r$, where mass loading begins.} for such systems will be 
\begin{equation}
R_{pi}=R_p\left(1-\frac{\Delta _r}{2 R_A}\right)
\end{equation} 
while the outer edge of the polar cap radius is  $\left(R_s/R_A\right)^{1/2}R_s$ \citep{poutanen09}, $R_s$ being the neutron star radius. For small values of $\Delta_r$ the columns would be hollow and thin walled. On the surface of the star this would create an accretion ring around the polar cap instead of a filled mound. To model such an accretion ring, we choose the mound height function to give a hollow mound in which the density falls off to zero both at the axis and at the polar cap radius. 

For the solution presented in Fig.~\ref{holoGS} we use a mound height profile as in eq.~(\ref{holopro}) with $Z_c = 45$m and $B_p = 10^{12}$G. The solution shows considerable distortion of field lines on both sides of the apex ($r \sim 698$m). This is in contrast to the case of filled mounds, where curvature of field lines occur towards the outer edge. Larger curvature of field lines allow larger mass to be accumulated per flux tube, as compared to that of filled mounds. Hence, although the central part is hollow, the total mass contained in the hollow mound ($M \sim 5.87\times10^{-13} M_{\odot}$), is comparable to that of a filled mound of the same height and field ( $M \sim 5.09\times10^{-13} M_{\odot}$  for $Z_c \sim 45$m and $B_p \sim 10^{12}$G and a parabolic profile as in eq.~\ref{parabolicpro}).   

The family of GS solutions for hollow mounds behave similarly as for filled mounds. With increase in maximum mound height $Z_c$, the GS solutions show larger curvature of field lines on both sides of ridge apex. The GS solutions fail to converge for mounds greater than a threshold height for a given magnetic field. For the mound height profile of eq.~(\ref{holopro}), the threshold height is around $Z_{\rm max} \sim 47$m for a polar magnetic field $B_p = 10^{12}$G. 
\begin{figure}
	\centering
	\includegraphics[width = 8cm, height = 8cm,keepaspectratio] {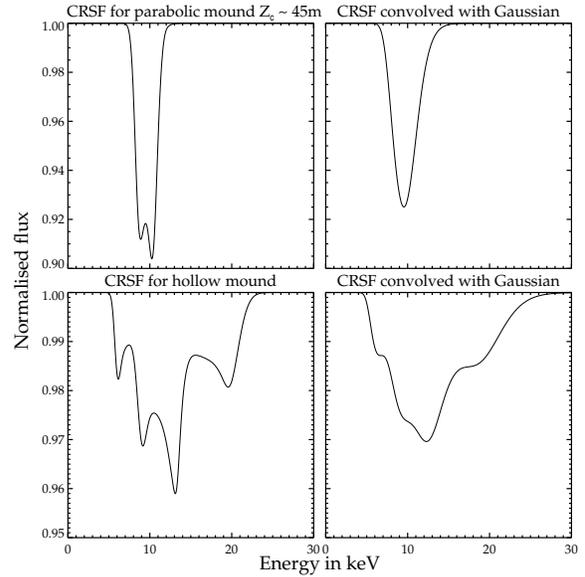}
	\caption{\small Top: CRSF from a filled mound of central height $Z_c \sim 45$m and $B_p \sim 10^{12}$G. The right panel gives the spectra convolved with a Gaussian (standard deviation 10\% of local energy) to simulate finite detector resolution. Bottom: CRSF from a hollow mound with $Z_c = 45$m and $B_p = 10^{12}$G, with the right panel giving the convolved spectra as before. The CRSF from hollow mounds show a much broader spectra due to contribution from different parts of the mound with large variations in the magnetic field. }\label{jointCRSF}
\end{figure}
\begin{figure*}
	\centering
	\includegraphics[width = 12cm, height = 6cm,keepaspectratio] {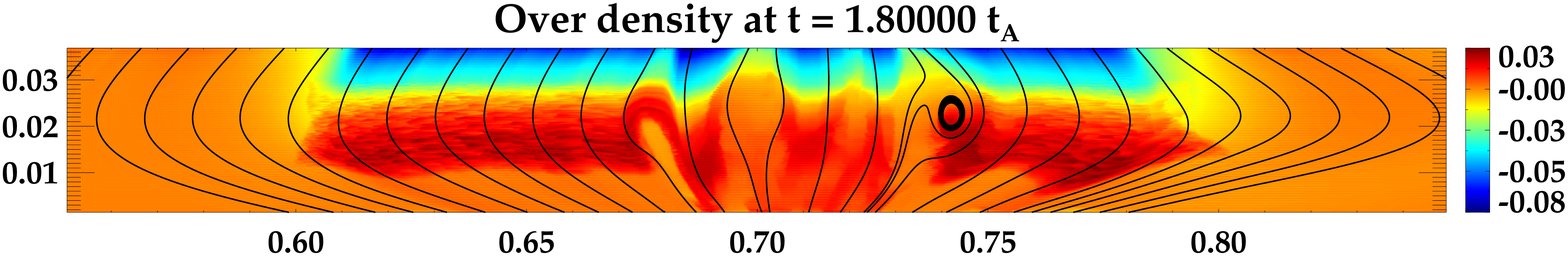}
	\includegraphics[width = 12cm, height = 6cm,keepaspectratio] {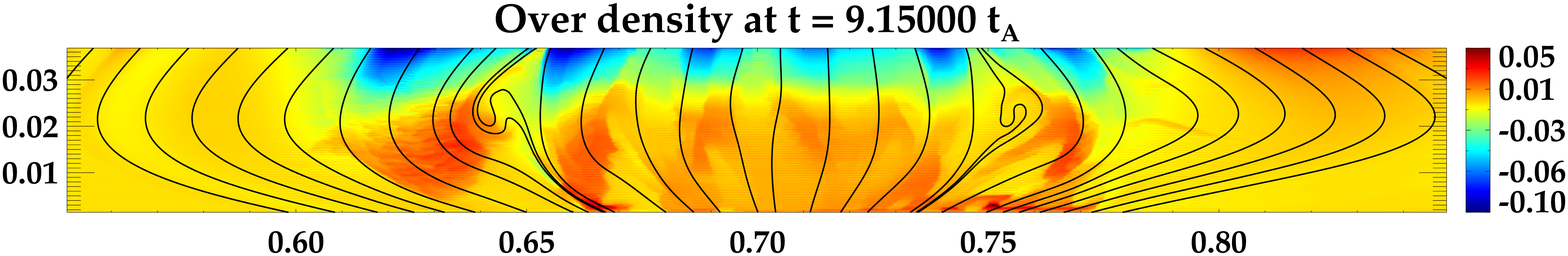}
	\includegraphics[width = 12cm, height = 6cm,keepaspectratio] {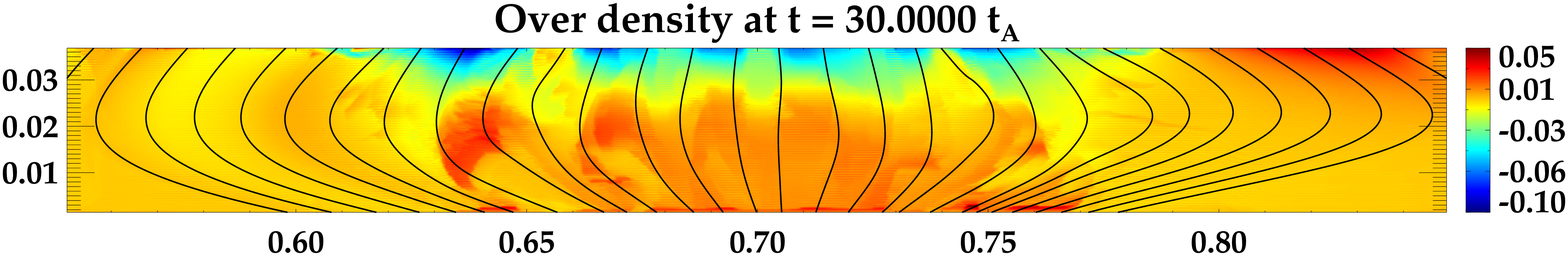}
	\caption{ \small Over density $(\rho - \rho _{\rm eq})/\rho_{\rm eq}$ and field line for hollow mound of maximum height $Z_c = 45$m and polar magnetic field $B_p = 10^{12}$G, with a positive density perturbation of strength $\eta = 5\%$. The simulation was carried out for a grid of size $1144\times 136$. The vertical and horizontal axes are the same as in Fig.~\ref{zeromean}. The perturbation results in formation of closed loops at multiple sites near the centre, very early in the simulation run.}\label{holoPLUTO_rho}
\end{figure*}
\begin{figure*}
	\centering
	\includegraphics[width = 12cm, height = 6cm,keepaspectratio] {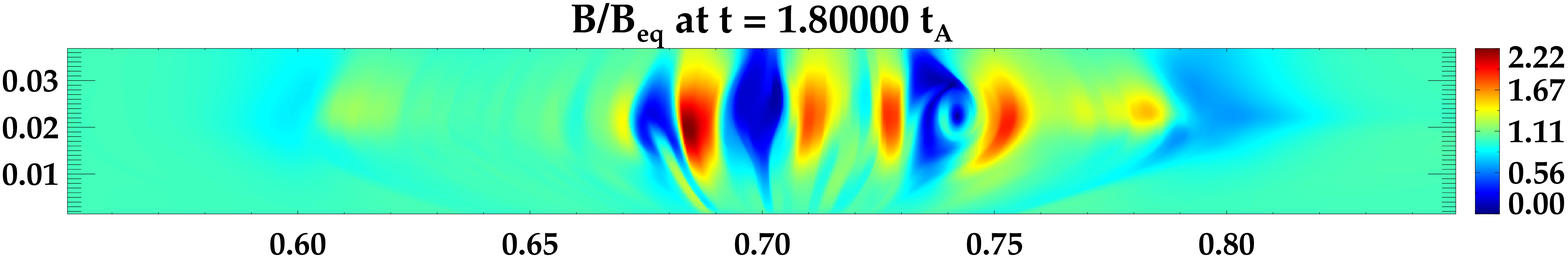}
	\includegraphics[width = 12cm, height = 6cm,keepaspectratio] {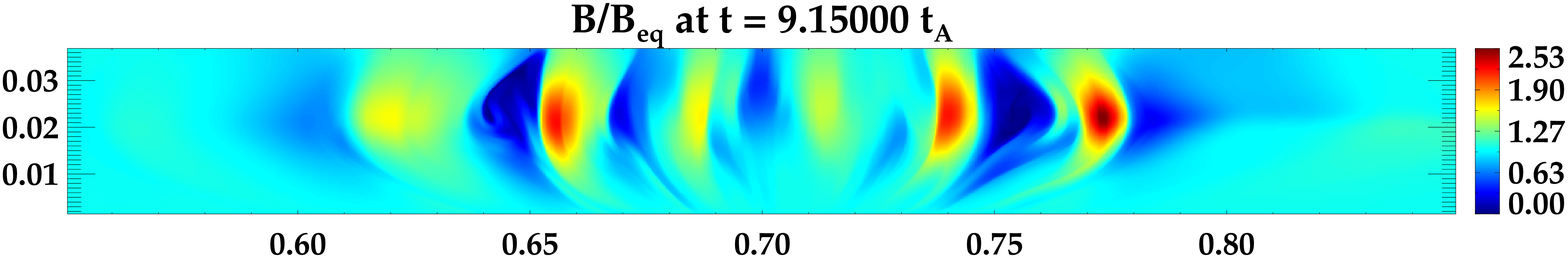}
	\includegraphics[width = 12cm, height = 6cm,keepaspectratio] {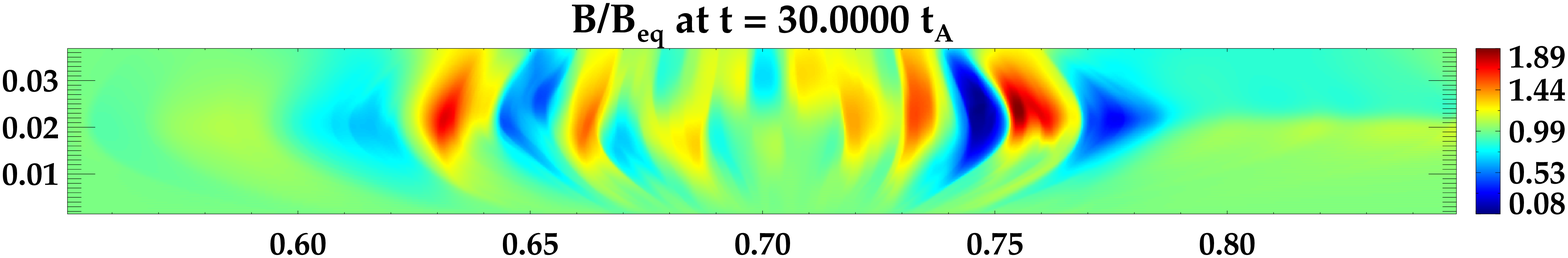}
	\caption{ \small  Magnetic field magnitude normalised to the local equilibrium value for the simulation described in Fig.\ref{holoPLUTO_rho}. Bunching of field lines in radial direction causes alternate regions of enhanced field strengths. The closed loops and pockets of enhanced fields migrate to the radial boundaries and eventually dissipate.}\label{holoPLUTO_bmag}
\end{figure*}
\subsection{Stability analysis of Hollow mounds}
Using the GS solutions for hollow mound, we perform stability analysis with PLUTO. The results are similar to that of a filled mound. Zero-mean density perturbations do not show growth of the perturbed region, indicating that the mounds are intrinsically stable with respect to interchange modes. For positive perturbations in density, closed loops are formed after a threshold perturbation strength. See Fig.~\ref{holoPLUTO_rho} and Fig.~\ref{holoPLUTO_bmag} for the results of a run with $\eta = 5\%$. The closed loops form quickly within a few Alfv\'en times and migrate away from the center, on both sides of the central height. This results in the formation of alternate regions of enhanced and reduced magnetic field due to the bunching of field lines, which have considerable departure from equilibrium solution. The field knots dissipate gradually as they migrate outwards.

\subsection{Cyclotron lines from hollow mounds}
Following the algorithm outlined in MB12, we have simulated the cyclotron resonance scattering features (hereafter CRSF) that will be observed in the emitted spectrum from a hollow mound. The spectra have been calculated by integrating the emission from different parts of the mound towards a given line of sight (hereafter los). We assume a Gaussian absorption profile whose depth and width are evaluated from interpolated results of \citet{schonherr07} for the slab 1-0 geometry. As in MB12, the line centre of the CRSF is obtained from the expression 
\begin{equation}\label{enform}
E_n = nE_{c0}\sqrt{1-u}\left(1-\frac{n}{2}\left(\frac{E_{c0}}{\rm 511 keV }\right)\sin ^2 \theta _{\alpha b}\right)
\end{equation}
where $n=1,2,3...$ is the order of the harmonic, $E_{c0}=11.6B_{12}$ in keV, $\theta _{\alpha b}$ is the angle between the direction of emission and local magnetic field and $u=r_s/r$, $r_s$ being the Schwarzschild radius. Emission from the inner part of the hollow mound may be blocked by the walls on the opposite side. In Appendix.~(\ref{shielding}) we explain the scheme we follow to account for such shielding. 

For the simulated spectra shown in Fig.~\ref{jointCRSF}, we consider emission from a single pole at inclination angle $\eta _p = 10^\circ$ and a los at $\i = 60^\circ$, both measured from the spin axis. The spectrum  shows multiple absorption features due to the large variation of field strength at the top of the mound (see Fig.~\ref{holofieldtop}). The different absorption features correspond to emission from different locations on top of the mound, with different magnetic field values.  The nature of this spectrum is significantly different from that expected from a filled parabolic mound of the same height (see Fig.~\ref{jointCRSF}). When convolved with a Gaussian of standard deviation $ \sim 10\%$ of the local energy, to simulate the finite resolution of a detector (see MB12 for details), the spectrum becomes a broad absorption feature. 

\section{Discussion and summary}\label{sec.disc}
\begin{figure}
	\centering
	\includegraphics[width = 9cm, height = 9cm,keepaspectratio] {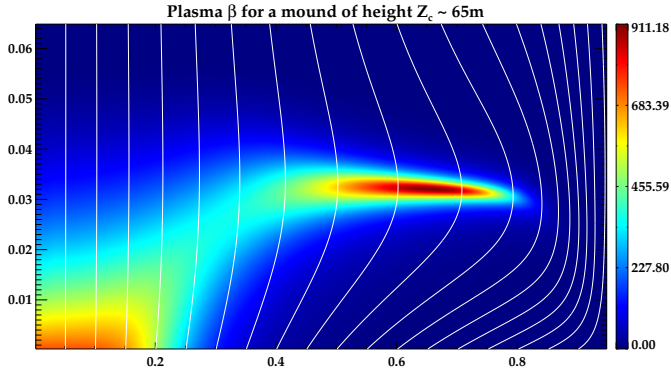}
	\caption{ \small Plasma $\beta$ (ratio of plasma pressure to magnetic pressure) for a GS solution of a  mound of height $Z_c \sim 65$m and $B_p \sim 10^{12}$G. The vertical and horizontal axes are the height and radius respectively, expressed in kilometres. The maximum plasma $\beta$ ($\sim 911$)  occurs along the central red horizontal patch near the regions of maximum curvature of the magnetic field lines (represented in white). At the regions of high $\beta$, the plasma is primarily supported by tension from curvature of field lines. Such regions are prone to pressure driven instabilities, and  show formation of closed loops when perturbed. }\label{beta}
\end{figure}
\begin{figure*}
	\centering
        \includegraphics[width = 16.0cm, height = 10cm,keepaspectratio] {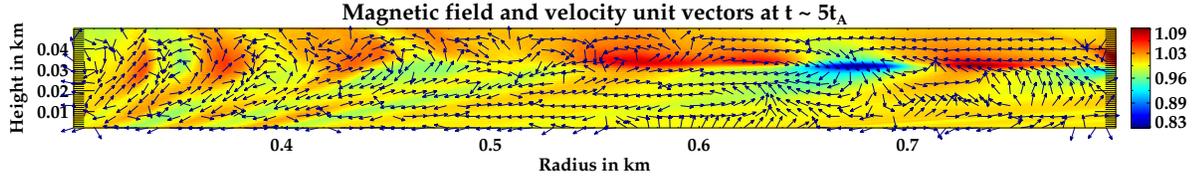}
	\caption{ \small Ratio of the magnetic field to its local equilibrium value for a barotropic simulation with random velocity perturbation of strength $\eta = 15\%$ of local sound speed (maximum initial velocity $\sim 0.35$, in normalised units). The velocity unit vectors are plotted to show the nature of the flow. Bunching of magnetic field takes place in the radial direction as local eddies are formed. The system settles down to a steady state with flow velocities less than $\sim 7.54\times 10^{-3}$ (in normalised units) at $t\sim5t_A$.}\label{velbaro}
\end{figure*}

\begin{enumerate}
\item
{\bf Absence of interchange mode instabilities: } In this paper we have tested for the stability of magneto-static accretion mounds by  MHD simulations using the PLUTO MHD code. From perturbation analysis we conclude that  mounds are stable with respect to interchange or ballooning modes in 2-D axisymmetric simulations\footnote{ Note that in this paper we consider a $T=0^\circ$K Fermi gas. However, finite plasma temperature can induce additional thermal modes \citep{cumming01} which have not been explored here.}. Linear stability analysis by \citet{litwin01} predict the onset of ballooning modes for a threshold plasma $\beta$ ($\beta = p/(B^2/8\pi)$). However such modes are inherently multi-dimensional in nature, with finite toroidal and zero poloidal wave vectors, normal to the local magnetic field (see \citet{friedberg82} for a review of MHD instabilities in confined plasma). Hence such modes cannot be excited in an axisymmetric 2D simulation.

 Litwin's approximate analytical estimates give a threshold $\beta _T \sim 11.7 (R_p/Z_c)$ for $\gamma \sim 5/3$, beyond which MHD instabilities will set in. For GS solution of a filled mound with $Z_c = 45$m and $B_p = 10^{12}$G, we get maximum $\beta \sim 293$, which is close to Litwin's threshold for the same mound $\beta _T \sim 260$. For higher mounds, $\beta _T$ decreases with increase in $Z_c$, and is much smaller than the maximum $\beta$ obtained from our GS solutions. For example, for a filled mound with $Z_c = 65$m and $B_p = 10^{12}$G, $\beta _T \sim 180$ whereas maximum $\beta \sim 911$ from the GS solution (see Fig~\ref{beta}).  Hence results from 2-D simulations cannot rule out the presence of such modes in a 3-D set up. Also, interchange mode instabilities \citep{chen84} can be excited in 3D simulation runs, as is seen in other examples of confined plasmas, e.g. in tokamak reactors. Work on 3-D stability analysis of accretion mounds is currently underway and will be addressed in a forthcoming publication (Mukherjee, Bhattacharya and Mignone in preparation).

\item
{\bf Instabilities due to excess mass: } From our 2-D simulations we have found that  addition of excess mass destabilizes the equilibrium due to gravity driven magnetic Rayleigh-Taylor type instabilities. For mounds with higher mass, the GS solutions have large radial (horizontal) component of magnetic field, which being perpendicular to gravity are also prone to Parker type instabilities \citep{cumming01,melatos01}. Topologically disconnected closed loops are formed beyond a threshold perturbation strength $\eta _T$.

From the expression of the energy integral for linear perturbations \citep{litwin01} on an adiabatic plasma ($p=k\rho^{\gamma}$), we have
\begin{equation}
\begin{aligned} \delta W &= \frac{1}{2} \int d^3x \{ \frac{\tilde{B}_\perp ^2}{4 \pi} + \frac{B^2}{4 \pi} \left(\boldsymbol{\nabla}\cdot\boldsymbol{\xi}_\perp + 2 \boldsymbol{\kappa _c}\cdot\boldsymbol{\xi}_\perp\right)^2 \\ 
&+ \gamma p\left(\boldsymbol{\nabla}\cdot\boldsymbol{\xi}-2\boldsymbol{\kappa}_g\cdot\boldsymbol{\xi}\right)^2  \\
 &- 2(\boldsymbol{\kappa}_c+\boldsymbol{\nabla}\phi/(2 c_s^2))\cdot\boldsymbol{\xi_\perp} \, (\boldsymbol{\nabla}p + \rho \boldsymbol{\nabla}\phi)\cdot\boldsymbol{\xi}_\perp \}
\end{aligned}
\end{equation}
where $\boldsymbol{\xi}$ is the plasma displacement, $\mathbf{\tilde{B}}=\boldsymbol{\nabla} \times (\boldsymbol{\xi}\times \mathbf{B})$ is the perturbed magnetic field, $\boldsymbol{\kappa}_c= (\mathbf{b}\cdot\nabla)\mathbf{b}$ is the magnetic field curvature vector, $c_s$ is the sound speed and $\phi$ the gravitational potential. $B_\phi$ is zero for our case.  

Instabilities will develop if the negative contribution from any (or all) of the terms containing field curvature, pressure gradient and gravity overcomes the stabilizing effects of the magnetic and pressure compression terms. \footnote{Necessary and sufficient condition for instability is: $\delta w <0$ \citep{bernstein58}.}  Hence, it is not a surprise that the closed loops are formed in regions with the largest curvature in field lines. This also corresponds to the regions with high plasma $\beta$, e.g. the red region in the middle of Fig.~\ref{beta} where $\beta \sim 911$. Pressure driven instabilities typically lead to a threshold plasma $\beta$ beyond which instabilities are triggered (e.g. \citet{friedberg82}, \citet{litwin01}). For mounds near the stability threshold, e.g. $Z_c \sim 72$m at $B_p = 10^{12}$G, the maximum plasma $\beta$ is as high as $ \sim 1.26 \times 10^4$.

The magnitude of $\eta _T$ decreases with increase in mound height, with $\eta _T \rightarrow 0$ as $Z_c \rightarrow Z_{\rm max}$, indicating inherent unstable nature of the mound for the modes under investigation. This corroborates the result of MB12 that GS solutions do not converge beyond a threshold height. The tests involving addition of mass are not meant to reflect realistic accretion rates. Although the amount of excess mass added in our simulations is small ($\sim 7.6\times10^{-15}M_\odot$ for $\eta=5\%$ perturbation on 65m mound), in a real system such mass will be accumulated slowly as mounds of larger mass are built. Effects of such inflow of material on an initially static mound have not been addressed here. However, from our current 2-D simulations we conclude that for large mound masses,  gravity and pressure driven modes result in the onset of MHD instabilities and no static equilibrium solution can be found beyond a threshold $Z_{\rm max}$. 

Buoyancy related instabilities due to the formation of topologically disconnected closed loops have previously been reported in the static mound simulations of \citet{hameury83} and \citet{melatos04}, and also dynamic MHD simulations by \citet{vigelius08} (hereafter VM08). However the threshold mass of the mound for the formation of closed loops in PM04 and VM08 is $M \sim 10^{-5}M_{\odot}$, which is much larger than the mass of the mounds in our present work. This may be due to the following differences in approach:
\begin{enumerate}
\item
PM04 and VM08 in their treatment consider spherical polar geometry and populate all field lines up to the equator, whereas we confine the accretion mounds strictly within the polar cap radius. Populating all field lines up to the equator provides lateral pressure support to the polar mound which can then hold a larger mass. 
\item
Plasma pressure due to isothermal EOS by PM04 and VM08  is several orders of magnitude less than the degenerate Fermi pressure used in our treatment, which results in higher plasma $\beta$ in our simulation. Such a system is more prone to pressure driven MHD instabilities e.g. \citet{friedberg82}. 
\end{enumerate}

\item
{\bf Adiabatic vs barotropic:}
We have also performed barotropic simulations with PLUTO for which the energy equation becomes redundant as pressure is evaluated from $p=k\rho^\gamma$, with $k$ a constant.  This is similar to the isothermal set up of MHD simulations. Results from adiabatic and barotropic modes are similar when perturbations are applied to velocity and magnetic fields. See Fig.~\ref{velbaro} for the results of velocity perturbation with barotropic simulation ($\eta = 15\%$ of local sound speed). The magnetic field bunches in the radial direction and local eddies are set up. The system settles down to a steady state with flow velocities reduced by more than 3 orders of magnitude at $t\sim 5t_A$. Similar results are also obtained for adiabatic EOS.  

However density perturbations behave differently in barotropic and adiabatic simulations. For a barotropic simulation, positive density perturbations on an initial static equilibrium create regions of excess pressure. The perturbed regions with high local pressure overcome the downward gravitational force and are quickly transported vertically upwards. Hence to study the effect of gravity driven modes due to the descent of added matter, adiabatic simulations have been performed in this work.

\item
{\bf Hollow mounds - structure and stability:}
We have solved the GS equation for mounds with hollow interiors. The hollow mounds show considerable distortion of the magnetic field on both sides of the maximum height to support the confined matter. There is a decrease in field near the ridge apex as field lines are pushed to either side. Closed loops form when excess mass is added to the equilibrium solution. The closed loops migrate to either side and eventually dissipate.

The fixed gradient boundary condition can induce artificial stability as it results in line tying type boundaries, which are known to give extra stability. In a real system, the plasmoids will be eventually ejected from the system. Plasma travelling inwards on closed loops may then eventually fill up the hollow. However there was no significant mass loss seen in our 2-D simulations.

\item
{\bf Hollow mounds - CRSF:}
CRSF from hollow mounds have been explored. From the simulation of the spectra integrated over the entire mound we see that: 
\begin{itemize}
\item
Cyclotron emission from the top of hollow mounds show complex fundamental features in the line shape (harmonics have not been evaluated), due to the large variations in magnetic field on top of the such mounds. This is similar to what is observed in the spectra of V0332+53  \citep{mowlavi06,nakajima10} which is conjectured to have a hollow column geometry \citep{ferrigno11}. Complex line shapes have also been predicted previously for strong non-dipolar local magnetic field by \citet{nishimura08,nishimura11}.
\item 
Convolving the CRSF with a Gaussian to account for finite energy resolution of detectors, we see that the resultant CRSF has the structure of a broad absorption envelope. 
\end{itemize}
Thus CRSF from hollow mounds will be characterised by broad line widths and complex structures in the line shape, which may be observed with improved detector resolution.

\end{enumerate}

Thus we conclude from this work that accretion mounds on neutron stars in HMXB are stable up to a threshold height and mass, beyond which MHD instabilities will disrupt the equilibria. Structure and stability of hollow mounds have been explored. It is shown that CRSF from such mounds will be characterised by broad features with a complex line shape. More work needs to be done to explore the 3-D stability of such systems and the effect of non-axisymmetric modes on the field structure and cyclotron emission from such mounds.

\section{Acknowledgement}
We thank CSIR India for Junior Research fellow grant, award no 09/545(0034)/2009-EMR-I. We thank Dr. Petros Tzeferacos for his help and suggestions in setting up the boundary conditions in the PLUTO simulations. We also thank Dr. Kandaswamy Subramanian, Dr. Ranjeev Misra and Sandeep Kumar from IUCAA, for useful discussions and suggestions during the work, and IUCAA HPC team for their help in using the IUCAA HPC where most of the numerical computations were carried out. We also the thank the anonymous referee for the detailed comments which have greatly helped in improving the work. DB acknowledges the hospitality of ISSI, Berne and discussions with the Magnet collaboration which have benefited the paper.

\appendix
\section{Shielding of radiation from inner walls of hollow mounds}\label{shielding}
\begin{figure}
	\centering
	\includegraphics[width = 8cm, height = 8cm,keepaspectratio] {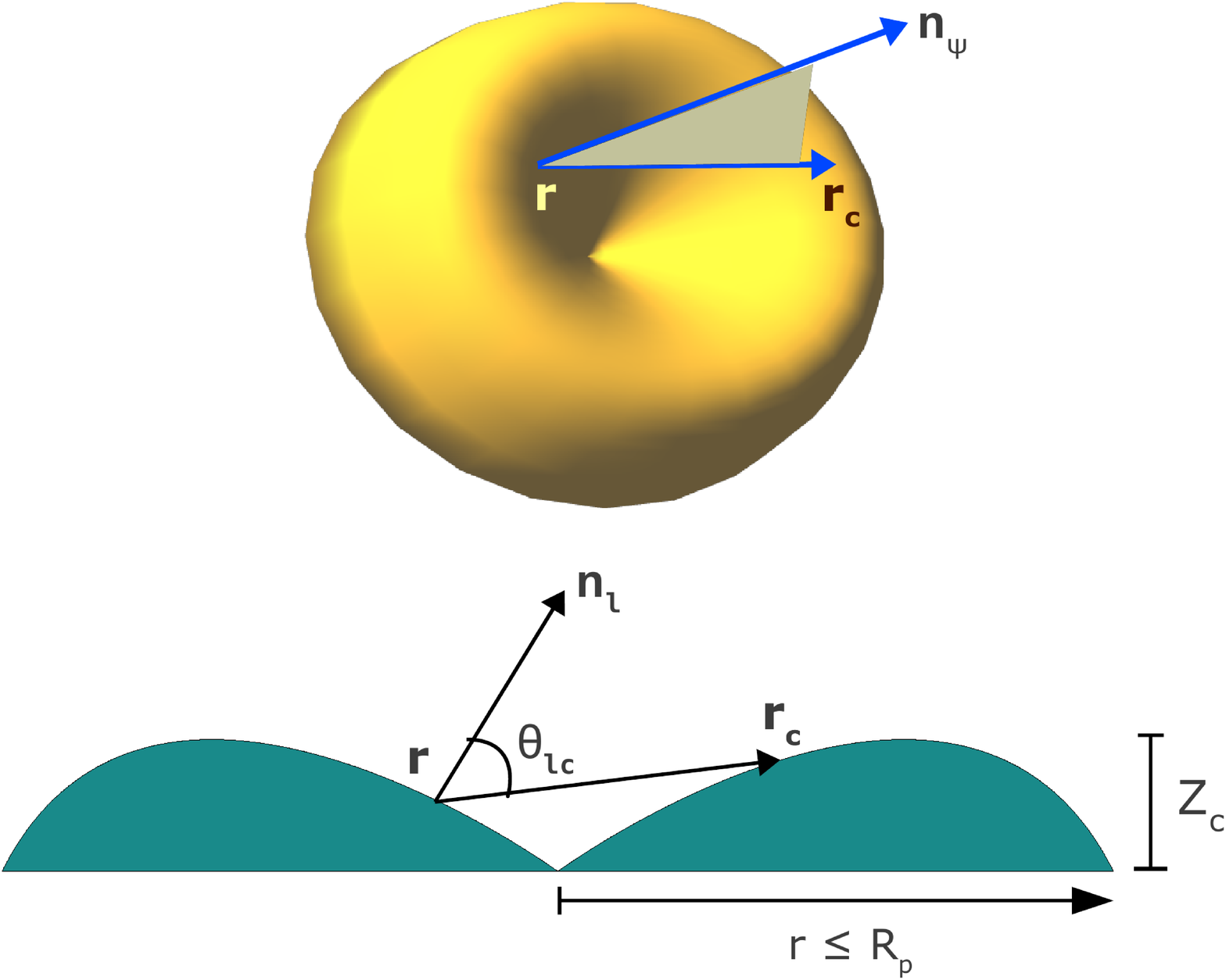}
	\caption{ \small Top: A 3-D schematic representation of the hollow mound. The vectors $\mathbf{n}_\psi$ and $\mathbf{r}_c - \mathbf{r}$ denote the plane where the path of the emitted ray to the observer lie. Bottom: a cross section of the mound along the plane of the emitted ray to the observer, and the location where the plane cuts the mound on the opposite side. }\label{schematicholo}
\end{figure}
In HMXBs an accretion column is formed by the infalling matter after it passes through a shock which may be several kilometres from the surface of the star, depending on the accretion rate (e.g. \citet{basko76}, \citet{becker07}, \citet{becker12}). In this work we consider the spectra generated from the mound without incorporating the effects of scattering from the overlying accretion column. This is valid for systems with low accretion rates and optically thin columns. The emission from the mound will then be directly visible and effects of overlying column will be small. However for systems with optically thick columns and large accretion rates, the emission from the mound will be obscured by scattering and absorption in the column. A proper Monte-Carlo simulation of the radiative transfer through the column must be carried out to address such cases, which will be reported in a future work (Kumar, Bhattacharya and Mukherjee in preparation). 

The rays of light coming from the hollow region can be blocked by the inner walls of the mound on the opposite side. Such rays will not contribute to the total spectra. The path of the emitted ray lies in the plane defined by the radius vector from the origin (centre of the neutron star) to the point of emission ($\mathbf{r}$) and the unit vector along the line of sight $\hat{\mathbf{n}}_\psi$ (see for e.g. \citet{beloborodov02}, \citet{poutanen06} and \citet{dipanjan12}). To exclude rays that may be blocked by the inner walls of the hollow mound, we first find the point where the plane defined by $\mathbf{r}$ and $\hat{\mathbf{n}}_\psi$ passes through the top of the mound $\mathbf{r}_c$ as in Fig.~\ref{schematicholo}. The radial and vertical coordinate of $\mathbf{r}_c$ ($r_c$ and $z_c$ respectively) are found by fitting a polynomial to the top of the mound obtained from the GS solution and evaluating the coordinate where $z$ is maximum. Since the three vectors $\mathbf{r}$, $\hat{\mathbf{n}}_\psi$ and $\mathbf{r}_c$ lie in the same plane, the angular coordinate $\phi _c$ of $\mathbf{r}_c$ is found from the condition 
\begin{equation}\label{eqphic1}
\mathbf{r}_c \cdot \left(\mathbf{r} \times \hat{\mathbf{n}}_\psi \right) = \mathbf{0}
\end{equation}
Following MB12 we use the following definitions for the vectors  $\hat{\mathbf{n}}_\psi \equiv (n_{\psi x}, n_{\psi y}, n_{\psi z}) \equiv (\sin i \sin \omega, \sin i \cos \omega, \cos i$) and  $\mathbf{r} \equiv (x,y,z) \equiv \lbrace \rho \cos \phi,\rho \cos \eta_p \sin \phi  + (\xi + R_s) \sin \eta_p,(\xi + R_s) \cos \eta_p  - \rho \sin \eta_p \sin \phi \rbrace$  where $i$ is the azimuthal angle of the observer's line of sight with respect to the spin axis, $\omega$ is the spin-phase angle, $(\rho,\phi,\xi)$ are coordinates of the emitting region in the polar cap frame with cylindrical coordinate system, $R_s$ is the neutron star radius and $\eta _p$ is the azimuthal angle of the centre of the polar cap. Using the above, we can rewrite eq.~(\ref{eqphic1}) as 
\begin{equation}\label{eqphic2}
A_c \cos \phi _c + B_c \sin \phi _c + C_c = 0
\end{equation}
where
\begin{eqnarray*}
A_c &=& \rho _c (y n_{\psi z} - z n_{\psi y}) \\
B_c &=& \rho _c \cos \eta _p (z n_{\psi x} - x n_{\psi z}) - \rho _c \sin \eta _p (x n_{\psi y} -y n_{\psi x}) \\
C_c &=& (\xi _c + R_s)\lbrace \sin \eta _p(z n_{\psi x} - x n_{\psi z}) \\
&&+ \cos \eta _p (x n_{\psi y} - y n_{\psi x}) \rbrace 
\end{eqnarray*}
Eq.~\ref{eqphic2} is solved using a modified Newton-Raphson scheme following \citet{press}. After finding the coordinate of $\mathbf{r}_c$ we evaluate the angle $\theta _{lc}$ (see Fig.~\ref{schematicholo}) between the local normal ($\hat{\mathbf{n}}_l$) and the radius vector from the point of emission to the top of the mound on the other side. 
\begin{equation}
\cos \theta _{lc} = \hat{\mathbf{n}}_l \cdot \frac{(\mathbf{r}_c - \mathbf{r})}{| \mathbf{r}_c - \mathbf{r} |}
\end{equation}
The normal vector is found as outlined in MB12 by evaluating the slope $ m_s=d\xi _{\rm top}/d\rho$ of the function $\xi_{\rm top}=f(\rho)$ ($\rho$ being the radial coordinate) that fits the top profile of the mound obtained from the GS solutions: $\hat{\mathbf{n}}_l \equiv \lbrace -\sin \theta _s \cos \phi, -\sin \theta _s \cos \eta_p \sin \phi  + \cos \theta _s \sin \eta_p, \cos \theta _s \cos \eta_p  +  \sin \theta _s \sin \eta_p \sin \phi \rbrace$, where $\sin \theta _s = \frac{m_s}{\sqrt{1+m_s^2}}$ and $\cos \theta _s = \frac{1}{\sqrt{1+m_s^2}}$. Using the above definitions of the vectors, one can write 
\begin{eqnarray*}
 \hat{\mathbf{n}}_l \cdot (\mathbf{r}_c - \mathbf{r}) &=& \cos \theta _s(\xi _c - \xi) + \sin \theta _s (\rho \cos \phi -\rho _c \cos \phi _c)\\
&& \times(\sin \phi + \cos \phi) \\
| \mathbf{r}_c - \mathbf{r} |^2 &=& \rho ^2 + \rho _c^2 +(\xi - \xi _c)^2 \\
&& - 2 \rho \rho _c (\cos \phi \cos \phi _c + \sin \phi \sin \phi _c)
\end{eqnarray*} 
 Any ray with emission angle larger than $\theta _{lc}$ will not contribute to the total spectra. This implicitly assumes that light will travel in a straight line and curvature effects from bending due to gravity is ignored for such short paths.

More accurate methods should be used to calculate the tangent vector from the point of emission to the mound surface on the other side. However, this involves more computation, and for sharp profiles of the hollow mound used, approximating the tangent point as the top of the mound will result in only a small correction. 

\def\apj{ApJ}%
\def\mnras{MNRAS}%
\def\aap{A\&A}%
\def\apjl{ApJ}
\def\physrep{PhR}
\def\apjs{ApJS}
\def\pasa{PASA}
\def\pasj{PASJ}
\def\nat{Natur}

\bibliographystyle{mn2e}
\bibliography{dipanjanbib}

\end{document}